\documentclass[aps,prl,showpacs,amsmath,amssymb,amsfonts,twocolumn,
superscriptaddress,floatfix,footinbib]{revtex4}

\usepackage{subfigure}
\usepackage{graphicx}
\usepackage[flushmargin]{footmisc}
\usepackage{color}
\usepackage{soul}
\usepackage{longtable}

\begin{document}

\title{Light Engineering of the Polariton Landscape in Semiconductor Microcavities}

  \author{A Amo}
    \affiliation{Laboratoire Kastler Brossel, Universit\'{e} Pierre et Marie Curie,
    \'{E}cole Normale Sup\'{e}rieure et CNRS, UPMC Case 74, 4 place Jussieu,
    75252 Paris Cedex 05, France}

  \author{S Pigeon}
    \affiliation{Laboratoire Mat\'{e}riaux et Ph\'{e}nom\`{e}nes Quantiques, UMR 7162,
    Universit\'{e} Paris Diderot-Paris 7 et CNRS, 75013 Paris, France}
  \author{C Adrados}
    \affiliation{Laboratoire Kastler Brossel, Universit\'{e} Pierre et Marie Curie,
    \'{E}cole Normale Sup\'{e}rieure et CNRS, UPMC Case 74, 4 place Jussieu,
    75252 Paris Cedex 05, France}
  \author{R Houdr\'{e}}
    \affiliation{Institut de Physique de la Mati\`{e}re Condens\'{e}e,
    Facult\'{e} des Sciences de Base, b\^{a}timent de Physique,
    Station 3, EPFL, CH 1015 Lausanne, Switzerland}
  \author{E Giacobino}
    \affiliation{Laboratoire Kastler Brossel, Universit\'{e} Pierre et Marie Curie,
    \'{E}cole Normale Sup\'{e}rieure et CNRS, UPMC Case 74, 4 place Jussieu,
    75252 Paris Cedex 05, France}
  \author{C Ciuti}
    \affiliation{Laboratoire Mat\'{e}riaux et Ph\'{e}nom\`{e}nes Quantiques, UMR 7162,
    Universit\'{e} Paris Diderot-Paris 7 et CNRS, 75013 Paris, France}
  \author{A Bramati}
    \affiliation{Laboratoire Kastler Brossel, Universit\'{e} Pierre et Marie Curie,
    \'{E}cole Normale Sup\'{e}rieure et CNRS, UPMC Case 74, 4 place Jussieu,
    75252 Paris Cedex 05, France}

%71.36.+c Polaritons (including photon-phonon and photon-magnon
% interactions)
%71.35.Gg Excitons and related phenomena, Exciton-mediated interactions 
%78.67.De Optical properties of low-dimensional, mesoscopic, and nanoscale materials and structures, Quantum wells

\date{\today}

\begin{abstract}

\noindent We demonstrate a method to create potential barriers with polarized light beams for polaritons in semiconductor microcavities. The form of the barriers
is engineered via the real space shape of a focalised beam on the sample. Their height can be determined by the visibility
of the scattering waves generated in a polariton fluid interacting with them.
This technique opens up the way to the creation of dynamical potentials and defects of any shape in semiconductor microcavities.
\end{abstract}

\maketitle
Optical beams have been used to trap and manipulate dielectric particles~\cite{Ashkin1970} and atoms~\cite{Chu1986}, as well as bacteria and intracellular organelles with nanometer resolution~\cite{Chu1991}. Optical cooling of atoms down to extremely low temperatures has also been achieved by controlling the momentum exchange between photons in a laser field slightly detuned from an atomic resonance~\cite{Cohen-Tannoudji1990}, giving access to the creation of atomic Bose-Einstein condensates (BEC)~\cite{Davis1995, Anderson1995}.
In atomic condensates, optical fields do not only allow for the cooling but also permit the engineering of the potential landscape seen by the condensate, taking advantage of weak light matter interactions, and have given rise to virtually any pre-designed configuration for the study of quantum fluids~\cite{IBloch2005}. For instance, an optical standing wave of the right energy is able to create periodic potentials whose minima act as deep traps for atomic gases. Combining standing waves in different directions has permitted the creation of BEC in two, one~\cite{Gorlitz2001} and zero dimensions, or the construction of random potentials~\cite{Billy2008}. One of the great advantages of this technique is that it allows for the dynamical modification of the potentials at high speeds. In this way, condensates can be stirred giving rise to the formation of vortex lattices~\cite{Abo-Shaeer2001}, and superfluidity can be studied by generating controlled velocity perturbations~\cite{Onofrio2000}. Condensates can also be dynamically divided resulting in the generation of squeezing and entanglement~\cite{Esteve2008}.

In the solid state sustems, optically induced traps have been demonstrated for indirect excitons~\cite{Hammack2006}, but this kind of potentials have not been so far used in microcavities, where polariton condensation has been observed~\cite{Kasprzak2006,Balili2007}. In this system, confining potentials have been created via partial or complete etching of microcavity samples during or after the growth, giving rise to samples of controlled dimensionality~\cite{Dasbach2002b,Kaitouni2006,Bajoni2008}. Another approach has been the deposition of thin metal stripes on top of an already grown planar microcavity~\cite{Lai2007b}, resulting in a blue shift of the photonic modes of up to $\sim400~\mu$eV ($\sim200~\mu$eV polariton shift at \emph{zero} cavity-exciton detuning). Both methods rely on structures with fixed designs preventing any post-processing manipulation. Potentials in microcavities have also been realized by means of pressure induced traps (up to 3-4~meV)~\cite{Balili2007} whose location can be varied, but they present a limited dynamic responses. Surface acoustic waves have also been used~\cite{deLima2006}, with configurations limited to undulatory periodic potentials. In this letter we present a direct all-optical method for the generation of potential barriers in semiconductor microcavities. Our technique is based on the blueshift induced by the polariton-polariton interactions in a high density polariton population, with a spatial design given by the shape of a control excitation laser. We show that polaritons created in the sample at lower densities strongly feel these barriers, which amount up to 1.5~meV in our experiments, showing strong scattering. Additionally, using a combination of polarization sensitive excitation and detection we can fully eliminate the transmitted light from the control beam, resulting in the observation of the signal polaritons of interest. This technique can be empowered with the use of currently available spatial light modulators in combination with pulsed lasers, giving access to a large number of potential configurations for the study of quantum phases in polariton condensates.

Our experiments have been performed in a $2\lambda$, GaAs/AlAs microcavity
with one In$_{0.04}$Ga$_{0.96}$As quantum well at each of the three antinodes of the confined electromagnetic field, with front/back reflectors with 21/24 pairs~\cite{Houdre2000}. The measured Rabi splitting at low temperature is 5.1~meV and we work at 5~K in a point of the sample with zero exciton-photon detuning.
Real and momentum space images of the emission in transmission geometry are collected using two high-definition CCD cameras.

\begin{figure}[t]
        \centering
        \includegraphics[width=0.75\columnwidth]{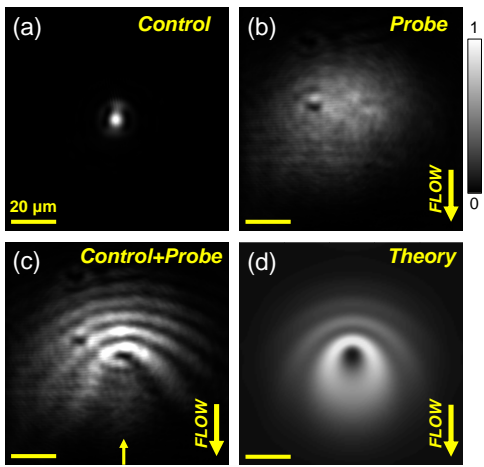}
    \caption{ Experimental real space emission of: (a)~a~point like potential generated by a control $\sigma^{-}$ laser (detection: $\sigma^{-}$),
    (b)~a~$\sigma^{+}$ probe polariton fluid in the linear regime (detection: $\sigma^{+}$), and (c) both the control potential and the probe fluid (detection: $\sigma^{+}$)~\cite{noteLightEnginnereing}. (d)~Image obtained from the solution of the Gross-Pitaevskii equation corresponding to (c)~\cite{noteLightEnginnereing}.}
    \label{fig:pointDefect}
    \end{figure}

In our experiments we use two cw excitation beams coming from the same laser, both with the same energy and close to resonant states of the lower polariton branch (LPB). The first one is a control beam which generates the engineered potential, while the second one is a probe beam which excites polaritons that interact with the induced potential. Figure~\ref{fig:pointDefect}a shows the real space image of a strong control beam focused in a tight gaussian spot of 4$\mu$m in diameter, and a wavelength of 837.08nm, blue-detuned by 0.1~meV from the emission of the LPB states with in-plane momentum $k=0$. If the density of excited polaritons is large, polariton-polariton interactions (arising from their exciton component) result in an appreciable blue-shift of the polariton energy over the area pumped by the control, given by $\Delta E=\hbar g\left|\psi\right|^2$, where $g$ is the polariton-polariton interaction constant and $\left|\psi\right|^2$ is the polariton density.
The probe beam excites an area of 45~$\mu$m in diameter (Fig.~\ref{fig:pointDefect}b) and it has an angle of incidence of 2.5$^\circ$ (in-plane momentum $k_p=0.33~\mu$m$^{-1}$). 

At low intensity of the probe field, polariton-polariton interactions are negligible, and do not give rise to any appreciable blueshift of the lower polariton branch energy. However, in the presence of the control beam (Fig.~\ref{fig:pointDefect}c), the probe polaritons experience a potential barrier in the spatial region where the high density control polaritons have induced a renormalization of the lower polariton branch. In this case, probe polaritons are scattered by the localized barrier induced by the control, giving rise to density waves. These waves are formed from the interference between the laser excited probe polaritons, in a plane wave, and the polaritons scattered in a cylindrical wave by the barrier. Their origin is analogous to the waves created by localized defects present in the sample~\cite{Carusotto2004,Amo2009,Amo2008} or by strong optical fields in moving atomic condensates~\cite{Carusotto2006}.

In order to clearly observe the effects of the induced barrier on the probe polaritons, we use different polarizations for the control and the probe beams. In Fig.~\ref{fig:pointDefect}, our beams are polarized in the following way: the control is circularly polarized $\sigma^{-}$ (giving rise to spin down polaritons), the probe is $\sigma^{+}$ (gas of spin up polaritons), and the detection is performed in the $\sigma^{+}$ polarized configuration. In this way only the probe polaritons are detected, preventing the saturation of the detectors by the strong control field.
Polariton-polariton interactions are strongly spin dependent~\cite{Renucci2005}, resulting in a larger effective interaction constant for polaritons with the same spin ($g_{\uparrow\uparrow}$) than with opposite spin ($g_{\uparrow\downarrow}$). For this reason, the renormalization of the LPB induced by the control polaritons is larger for polaritons of the same spin as those of the control, but the non-vanishing value of $g_{\uparrow\downarrow}$ results in an appreciable renormalization also for probe polaritons (of opposite spin) as evidenced in Fig.~\ref{fig:pointDefect}c.

    \begin{figure}[t]
        \centering
        \includegraphics[width=0.75\columnwidth]{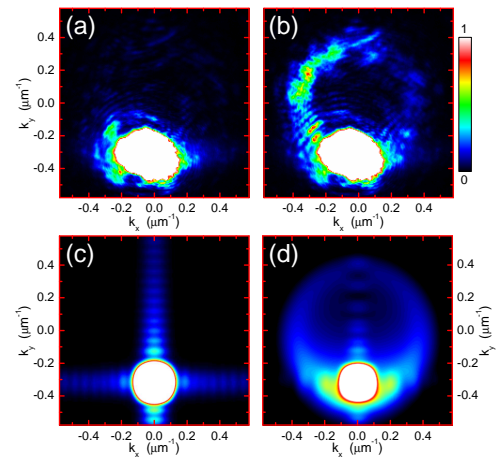}
    \caption{(color online) (a),(b) Experimental far field images in the absence (a) and in the presence (b) of the control beam, corresponding to the near field images in Fig.~\ref{fig:pointDefect}b,c, respectively, showing the scattering of probe polaritons generated by the induced potential barrier. The white saturated areas correspond to the transmitted probe. (c), (d) Simulated far field images corresponding to (a) and (b), respectively. Vertical and horizontal traces in (c) arise from the periodicity of the numerical spatial grid.}
    \label{fig:farField}
    \end{figure}

Figure~\ref{fig:pointDefect}d shows the real space image in the conditions of Fig.~\ref{fig:pointDefect}c, obtained from the numerical solution of the spin dependent Gross-Pitaevskii equation~\cite{Carusotto2004,Shelykh2006,Amo2008}, both images being in very good quantitative agreement. In the simulations we have used a value of $g_{\uparrow\uparrow}=6.2\times10^{-4}$~meV$ \mu$m$^2$~\cite{Ciuti1998b,Amo2008}. The experimental results are succesfully reproduced with $g_{\uparrow\downarrow}=+0.1g_{\uparrow\uparrow}$. Recent calculations show that the magnitude and sign of $g_{\uparrow\downarrow}$ depend on the polariton momentum, exciton-cavity detuning and bi-exciton energy~\cite{Wouters2007c}. The value of $g_{\uparrow\downarrow}$ has already been studied in some experiments~\cite{Renucci2005}. In our case we can evaluate its order of magnitude as given above. Further experiments should allow a more precise determination.

The scattering of probe polaritons with the induced barrier is also evidenced in the far-field of the emission. Figure~\ref{fig:farField} shows the experimental momentum distribution of the probe polaritons in the absence (a) and in the presence (b) of the control beam, corresponding to the real space images of Fig.~\ref{fig:pointDefect}b and c, respectively. The cilindrical scattering of probe polaritons on the barrier induced by the control beam gives rise to a significative Rayleigh ring in momentum space (Fig.~\ref{fig:farField}b). Figure~\ref{fig:farField}c,d show the calculated images corresponding to a,b, respectively, evidencing a good qualitative agreement.

    \begin{figure}
        \centering
        \includegraphics[width=0.8\columnwidth]{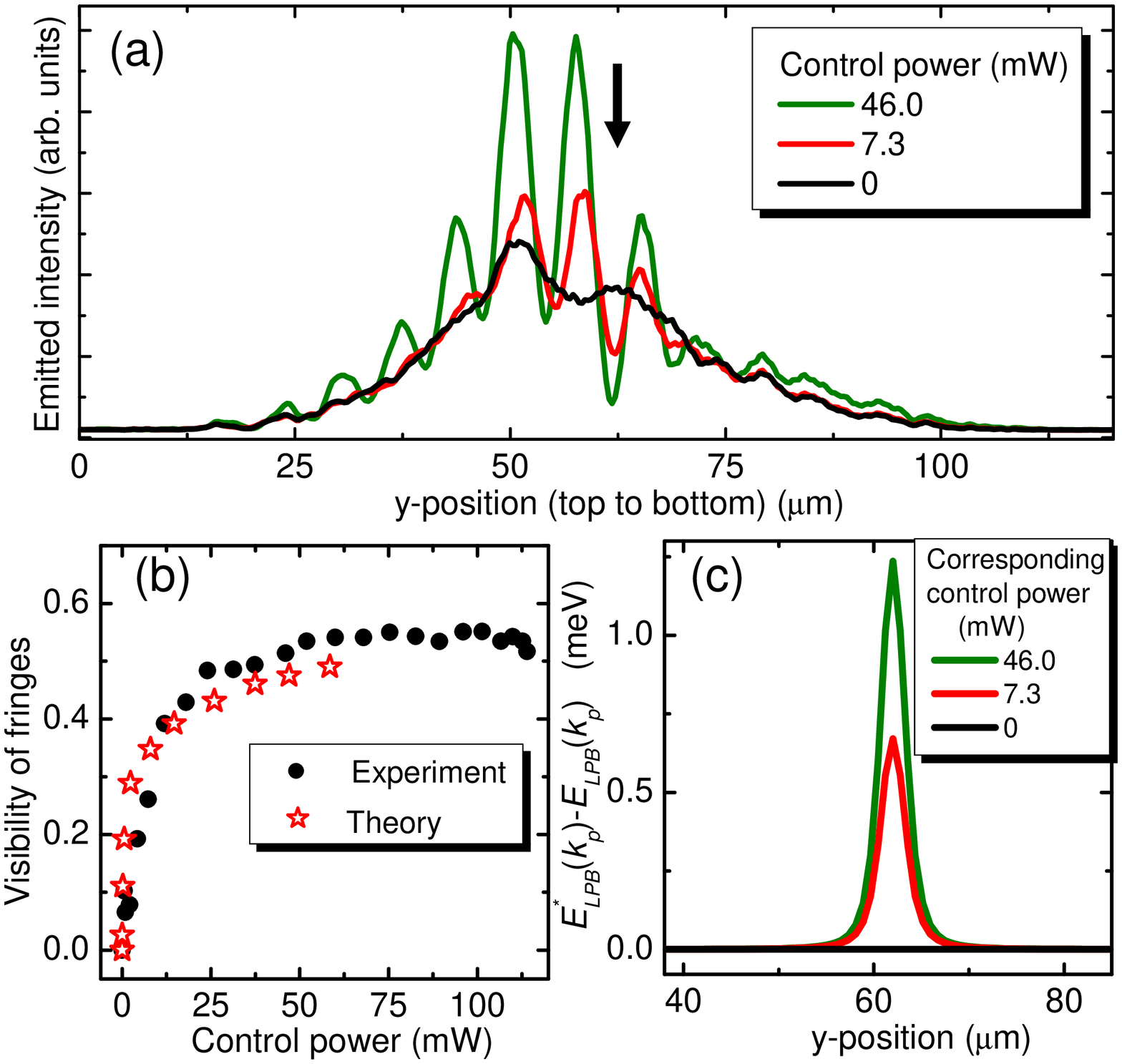}
    \caption{(color online) (a) Real space $y$-profiles along the direction indicated by the arrow in Fig.~\ref{fig:pointDefect}c for three powers of the control beam. The solid arrow indicates the $x$-position of the induced barrier. (b) Measured (solid points) and calculated (stars) visibility of the fringes as a function of the power of the control beam. The visibility of the fringes is obtained from the intensity of the first two maxima and minima behind the induced potential [to the left of the black arrow in (a)]. The corresponding real space images can be seen in~\cite{noteLightEnginnereing}. (c) Calculated height of the induced potential corresponding to the control powers depicted in (a).}
    \label{fig:visibilityFringes}
    \end{figure}

The height of the potential barrier induced by the control field can be directly tuned via its intensity. Figure~\ref{fig:visibilityFringes}a shows intensity profiles taken along a vertical cut across the direction indicated by the arrow in Fig.~\ref{fig:pointDefect}c, for different control intensities in the conditions of Fig.~\ref{fig:pointDefect}. As the power of the control is increased the visibility of the fringes increases (black dots in Fig.~\ref{fig:visibilityFringes}b), indicating that the polariton scattering on the induced potential is larger, a consequence of the increased potential barrier created by the control. This phenomenon is demonstrated by our calculations, which show a correlation between the calculated visibility of the fringes depicted by stars in Fig.~\ref{fig:visibilityFringes}b, and the calculated height of the induced barrier as a function of control power, depicted in Fig.~\ref{fig:visibilityFringes}c for the conditions of Fig.~\ref{fig:visibilityFringes}a. Note that induced renormalizations as large as $\sim$1.5~meV can be easily obtained.

    \begin{figure}
        \centering
        \includegraphics[width=\columnwidth]{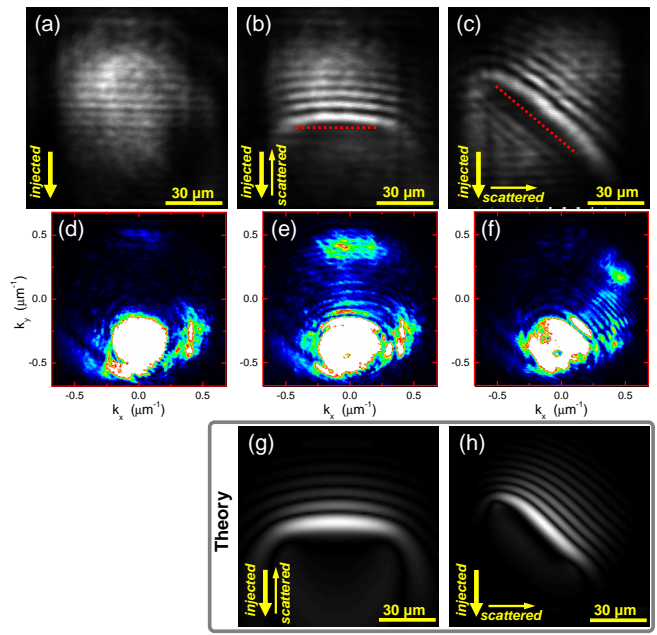}
    \caption{(color online) Real [momentum] space emitted intensity of a TM polarized probe in the linear regime, alone (a) [d] and in the presence of a line potential induced by a TE polarized control beam in the horizontal (b) [e] and diagonal directions (c) [f]. (g) and (h)~display simulated images corresponding to~(b) and (c), respectively. The dashed red lines indicate the orientation and position of the control. All images are detected along the TM polarization. Control and probe beams are detuned by 1.0~meV from each other.}
    \label{fig:lineDefect}
    \end{figure}

The flexibility of our technique allows us to explore the creation of potential barriers of different shapes in Fig.~\ref{fig:lineDefect}. In Fig.~\ref{fig:lineDefect}b we study the situation of a linear barrier placed perpendicular to the flow of probe polaritons and as wide as the probe spot. In this case, the control is linearly polarized transverse electric (TE) while the probe and the detection are transverse magnetic (TM) linearly polarized. In this case control and probe beams are created by two different laser, their photon energies being detuned from each other by 1.0~meV. The linearly polarized control injects polaritons with both spins, inducing the same renormalization for both polariton spinors. In this way, the effective potential barrier seen by the probe is larger than in the case of the circular polarization configuration depicted in Figs.~\ref{fig:pointDefect},~\ref{fig:farField} and~\ref{fig:visibilityFringes}, as control and probe polaritons interact directly via the $g_{\uparrow\uparrow}$ term. Qualitatively similar results were obtained with circularly polarized control and probe.

The induced barrier creates a strong scattering of the probe polaritons in the direction perpendicular to the barrier, resulting in the generation of linear density waves parallel to the barrier in the upstream direction, analogous to those observed in Fig.~\ref{fig:pointDefect}c for a point-like potential. In the far field (Fig.~\ref{fig:lineDefect}e), the barrier-induced retro-reflection of probe polaritons is manifested by the appearance of a peak with opposite momentum to that of the probe beam.

If the line-shaped control is placed with an inclination of 45$^\circ$ with respect to the probe flow (Fig.~\ref{fig:lineDefect}c), probe polaritons are reflected (scattered) by the induced barrier towards the horizontal direction. In this case the interference between the polaritons injected by the probe (flowing down in the figures) and the scattered polaritons results in waves whose maxima are oriented parallel to the direction of the control induced barrier. In the far field, the scattered polaritons give rise to a peak in a position close to $(k_x,k_y)\approx(0.4, 0.2)\mu$m$^{-1}$, as evidenced in Fig.~\ref{fig:lineDefect}f.
Figures~\ref{fig:lineDefect}g,h depict the results obtained by solving the spin-dependent Gross-Pitaevskii equation in the conditions of Fig.~\ref{fig:lineDefect}b,c (linear polarizations) showing again good quantitative agreement with the experimental data.

    \begin{figure}
        \centering
        \includegraphics[width=0.8\columnwidth]{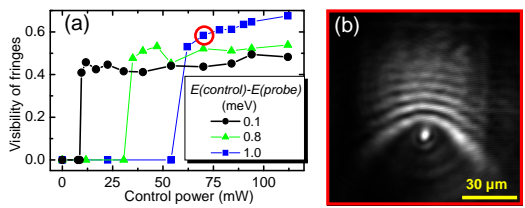}
    \caption{(color online) (a) Visibility of fringes as a function of control power for different control-probe detunings, for a point-like control, and same polarization conditions as in Fig.~\ref{fig:lineDefect}. (b) Probe emission in the conditions marked by a circle in (a).}
    \label{fig:detuning}
    \end{figure}

Note that in Fig.~\ref{fig:lineDefect}, the control and probe beams have different wavelengths. In Fig.~\ref{fig:detuning}a we study the scattering induced by a point like barrier for different control detunings with respect to the probe, under the same polarizations conditions as in Fig.~\ref{fig:lineDefect}, via the visibility of the generated fringes (Fig.~\ref{fig:detuning}b). Values of the visibility of the fringes similar to those shown in Fig.~\ref{fig:visibilityFringes}b are obtained above a threshold, which arises from the non-linear character of this
phenomenon~\cite{Ciuti2005}. When the control field is detuned from the LPB energy, injection of polaritons is only efficient
above a given threshold density, at which the LPB abruptly renormalizes up to the energy of the pump~\cite{Ciuti2005}. Larger
control beam detunings result in higher thresholds. In the configuration of non-degenerate control and probe fields, probe signal polaritons can be selected spectrally, without the need of the polarization selection used in our experiments.

Our results show the capability to tailor the potential landscape in semiconductor microcavities with the use light fields thanks to the strong polariton-polariton interactions. This is a crucial element for the study of quantum fluid effects in engineered potentials, for instance, in confined geometries. Optically induced barriers will enable the study of polariton Josephson oscillations~\cite{Sarchi2008} across an energy wall of tunable height, polariton trapping in light induced micropillars, or localization effects~\cite{Gurioli2005} in speckle generated random potentials~\cite{Billy2008}. Additionally, linear barriers as those described in Fig.~\ref{fig:farField} allow for the controlled scattering of polaritons into a pre-defined direction given by the shape and orientation of a control beam. This configuration presents interesting potential applications in the optically controlled multiplexing of light beams at high modulation rates (in the THz range, limited by the polariton lifetime).

This work was supported by the IFRAF and the \emph{Agence Nationale pour la Recherche}. A.B. is a member of the Institut Universitaire de France.

%-----%-----%
%\begin{thebibliography}{10}
%-----%-----%

%\bibliography{articulosalberto}
%\end{thebibliography}

\end{document}